# Estimation of Critical Temperature of High-Temperature Superconductors from AC susceptibility measurements using a pair of Neural Networks

Marcin Kowalik, Michał Kowalczyk, Jan M. Michalik, Marek Giebułtowski, Ryszard Zalecki, Janusz Niewolski and Waldemar Tokarz

Department of Solid State Physics, Faculty of Physics and Applied Computer Science,
AGH University of Science and Technology, A. Mickiewicza 30 Av., 30-059 Cracow, Poland

**Abstract**

This paper demonstrates the method of estimation of critical temperature $T_c$ value of high-temperature superconductors from the dispersive part of AC susceptibility measurement using a pair of neural networks.

## 1. Introduction

A superconductor is a material, which cooled below a certain temperature, called critical temperature $T_c$, has exactly zero electrical resistance. Below $T_c$ the expulsion of the external magnetic field from the inside of superconductor material is also observed and is called a Meissner effect. The critical temperature $T_c$ is the most basic characteristic of the superconducting material [1].

In 1986 the cuprate-perovskite ceramic material Ba-La-Cu-O with $T_c$ about 30 K was discovered [2]. A year later superconductivity transition between 80 K and 93 K was observed in the Y-Ba-Cu-O system [3]. Research in the following 30 years has led to the discovery of numerous cuprate superconductors that belong to several families with $T_c$ up to 134 K for $HgBa_2Ca_2Cu_3O_9$ at ambient pressure [4] and as high as 164 K under 30 GPa [5]. These ceramic materials with a high value of critical temperature became known as High-Temperature Superconductors (HTS).

The artificial neural network (NN) is modelled on a biological brain. It simulates the behaviour of many of densely interconnected artificial neurons, which can learn and make a decision through associations with training data.. NNs are not strictly programmed, they learn by themselves based on provided data (information) [6]. The most recent accomplishments in the area of HTS reported by Maddury Somayazulu et al [7] demonstrated that lanthanum superhydride at 180–200 GPa pressure display superconducting properties in temperatures as high as 250-260 K.

In recent years the neural networks (NN) demonstrated the ability to classify with a very high degree of accuracy sets of labelled data [8, 9]. Neural network with ease can properly recognize and classify handwritten digits or letters [10], phases in condensed matter physics [11] or even light curves of stars in searching for exoplanets [12]. NN can also be used to solve regression problems.

In this paper, we show the method of estimation critical temperature $T_c$ value of HTS from the dispersive part of AC susceptibility measurement (figure 1) using a pair of neural networks. Application of neural networks to this task should allow for precise estimation of value $T_c$ and other parameters, which can be evaluated directly from measurement data, even in cases where distinct noise or small but still apparent unphysical contributions from laboratory equipment are apparent. Estimation of $T_c$ analytically can be very difficult when taking into account these kinds of obstacles.

## 2. Theory and experimental details

The AC magnetic susceptibility can be described as a complex number by the formula $\chi=\chi'+i\chi''$, where $\chi'$ is the dispersion and $\chi''$ is the absorption part of the dynamic susceptibility. The value of the dispersion part corresponds to the diamagnetic response of the HTS sample when an external magnetic field is applied. The value of absorption part coincide to the energy converted into heat during one cycle of the external, AC magnetic field $H_{ac}$. For the bulk HTS samples this energy loss is connected with the magnetic field penetration into the intra and inter-granular regions [13].

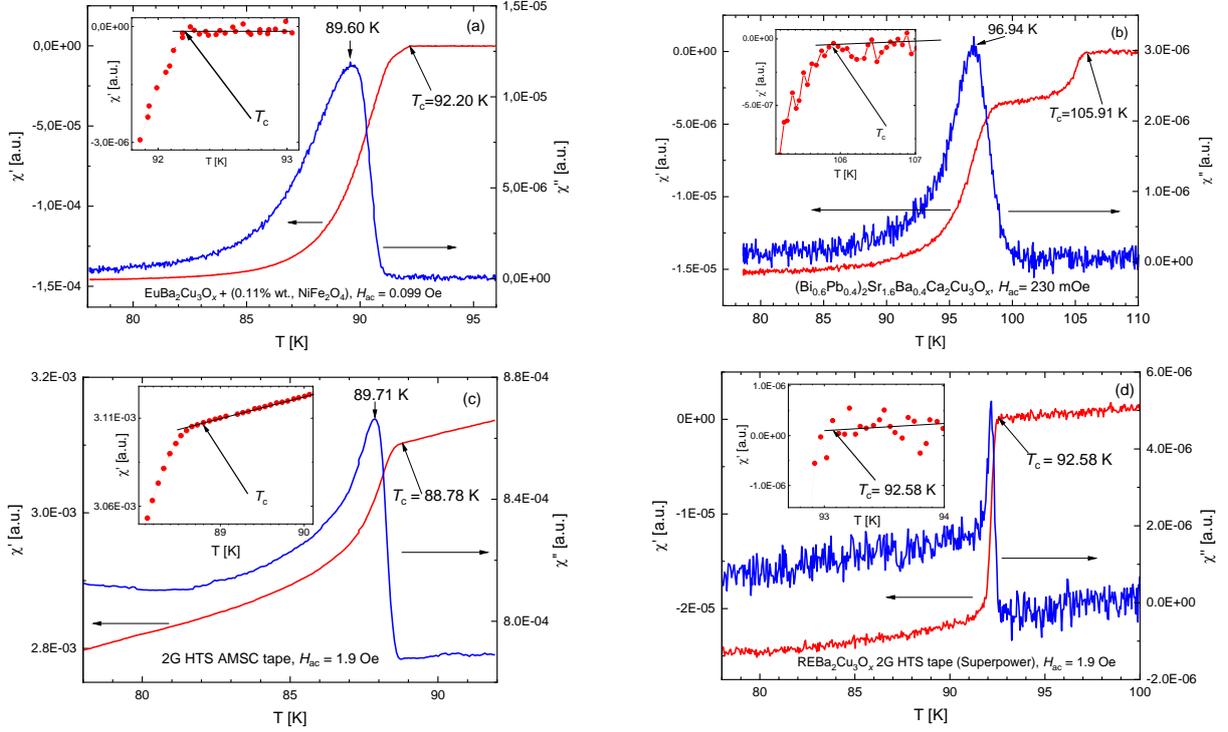

**Figure 1.** The dispersive χ' and absorption χ' part of AC susceptibility vs temperature for mixture of polycrystalline $EuBa_2Cu_3O_x$ HTS and 0.11% wt. $NiFe_2O_4$ nanoparticles (a), polycrystalline $(Bi_{0.6}Pb_{0.4})_2Sr_{1.6}Ba_{0.4}Ca_2Cu_3O_x$, (b), thin layer 2G HTS tape manufactured by American Superconductors (c) and thin layer $REBa_2Cu_3O_x$ 2G HTS tape manufactured by Superpower (d). The insets show $T_c$ values estimated by an expert.

The values of χ' and χ" for HTS change with temperature. In the ideal case, above a certain temperature, called the critical temperature $T_c$, the values of both parts of AC susceptibility are equal to zero. On the other hand, below critical temperature $T_c$, the χ' part has negative values and the χ" part is positive or equal to zero. These features can be used to define the $T_c$ of HTS [14]. Shapes of χ'(T) and χ"(T) curves depend on superconducting properties of the HTS sample i.e. species of HTS, microstructure (bulk, single grain, thin layer), oxygen parameter, preparation procedure. In the figure 1 χ(T) curves for selected species of HTS samples are shown.

NN's were trained on the dataset consisting of several hundred measurements of χ'(T) for different species of HTS. For every single measurement, the corresponding value of $T_c$ was determined by hand. The critical temperature $T_c$ was defined as the temperature at which the diamagnetic signal of the sample χ' appeared (i.e. χ' became negative), when the temperature of the sample was decreasing (see insets of figure 1).

The dataset contained results of measurements for polycrystalline HTS like: $YBa_2Cu_3O_x$ [15], $EuBa_2Cu_3O_x$ (REBCO-123, where RE is rare earth element), $(Bi_{1-y}Pb_y)_2Sr_2Ca_2Cu_3O_x$ (BISCO-2223), $Tl_2Ba_2Ca_2Cu_3O_x$ (Tl-2223) and thin layer 2G HTS tape manufactured by SuperPower Inc. Samples were synthetized by the solid state reaction in various conditions. The samples were characterised by XRD, SEM, EDX

Selected samples have been carefully examined using Scanning Electron Microscopy with Energy-Dispersive X-ray Spectroscopy (EDS) and magneto-resistance measurements. Secondary electron images allowed an observation of the microstructure of the samples (see figure 2). Backscattered electron imaging together with EDS analysis determined stoichiometry and homogeneity of the samples.

The AC susceptibility measurements were performed as a function of temperature and AC magnetic field $H_{ac}$ up to 10.9 Oe. Temperature range for REBCO samples was 77-100K and 77-125K for BISCO and Tl-2223.

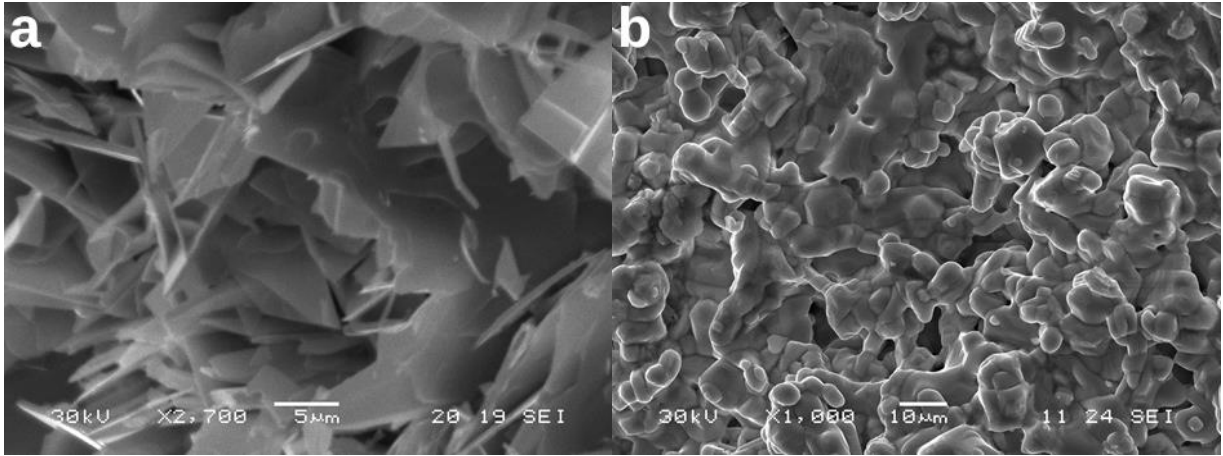

**Figure 2.** SEM micrographs showing of microstructures of BISCO-2223 (a) and $ErBa_2Cu_3O_x$ HTS (b).

The AC susceptibility spectrometer was based on three coils inductance bridge, which was placed in the helium cryostat [16]. A SRS830 DSP Lock-In amplifier was used as a detector and the AC current source at 189 Hz. The temperature was monitored by the Lake Shore Model 330 autotuning temperature controller employing chromel-gold – 0.07% Fe thermocouple with the accuracy of about 0.3 K and resolution of about 0.05 K. The control of the measurements and data acquisition was performed with a computer. The direction of an applied magnetic field $H_{AC}$ was parallel to the longest side of the parallelepipedal sample.

## 3. The method for automatic estimation of $T_c$

To evaluate the value of $T_c$ from single $\chi(T)$ measurement a pair of NN was needed. The goal of the first NN was to select by classification a few subsequences of $\chi'(T)$ curve (single measurement), which were useful for $T_c$ estimation. The goal of the second NN was to guess the value of $T_c$ using regression technique on a given subsequence.

A single measurement consisted of about 400 data points on average. A single data point- is a vector consisting of three values $\chi'$, $\chi''$ and the corresponding temperature of the sample. In each measurement, the minimal value of $\chi'(T)$ were normalized to -1. Subsequences were generated using the sliding window approach on normalized measurement [12]. Each subsequence is a vector with 60 subsequent values of $\chi'$, ordered by increasing temperature. The values of $\chi''$ and sample temperature were excluded from analysis by NNs based on the assumption that all data points are evenly spaced through temperature, which is used for simplification of analysis.

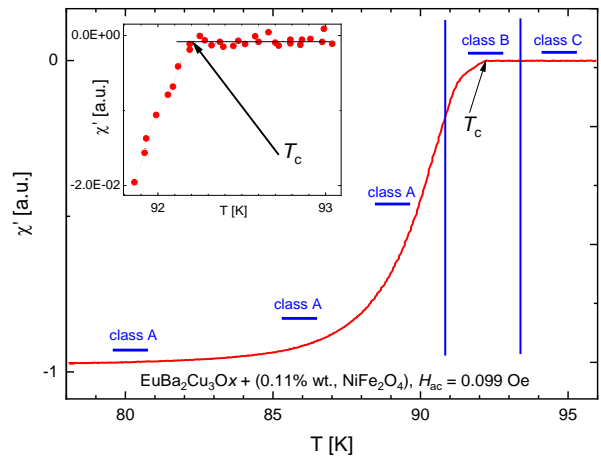

**Figure 3.** An exemplary division of the dispersive part of AC susceptibility (measuremed for a mixture of polycrystalline $EuBa_2Cu_3Ox$ HTS and 0.11% wt. $NiFe_2O_4$ nanoparticles) into subsequences belonging to one of three classes A, B, C. Short blue horizontal lines mark a single subsequence of 60 subsequent values of $\chi'$. The inset shows the value of $T_c$ estimated by an expert.

The classification NN had to learn the features of subsequences of χ'(T) curve and classify them into three distinct classes marked as class A, B and C (figure 3). The classification was based on the χ' values present in the subsequence. All values of χ'(T) were negative and noticeably smaller than zero in a class A. On the other hand, all values of χ'(T) in class C subsequence were very close to zero and were negative or positive. Finally, the class B subsequence had the first part of the values of χ'(T) noticeably smaller than 0 and the rest close to zero, so class B subsequences incorporates values from class A and C. The classification NN given a subsequence returns a three element vector, where the values are probabilities of subsequence belonging to class A, B or C. Thus the sequence of classification results will show probabilities of the three classes of subsequences as a function of temperature. An ideal case of classification is shown on figure 4a.

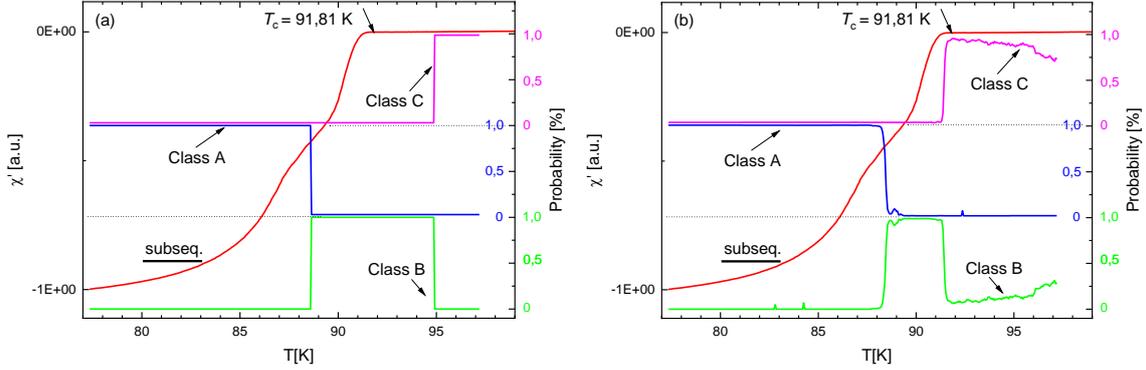

**Figure 4.** The probability distribution for an ideal case (a) and a **probability distribution calculated** by NN (b) of subsequence classification into one of three classes: A, B and C as a function of temperature. The short black line shows the length of subsequence consisting of 60 subsequent values of χ'.

The second NN estimated the value of $T_c$ using regression technique only on class B subsequences. The NN predicted the temperature distance (marked as $d$) between the unknown point, which determined the value of $T_c$, and the centre point of the subsequence (figure 5). The estimated value of $T_c$ could be calculated according to the formula: $T_c = T_1 + (1 + d)\frac{T_n - T_1}{2}$, where $T_1$ and $T_n$, are the sample temperatures of first and last data point of a given subsequence of class B.

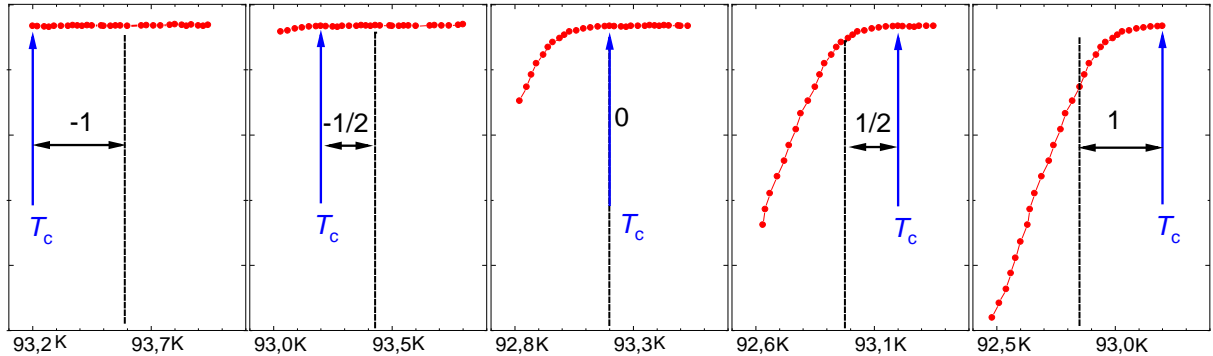

**Figure 5.** The method of $T_c$ estimation for different subsequences of a size of 30 data points. The point which determines the value of $T_c$ is indicated by the blue arrow. The values -1, -1/2, 0, 1/2, 1 show the normalised distance $d$ between the $T_c$ and centre of the subsequence, which is indicated as a dotted black line.

## 4. Neural networks architectures

The classification and regression NNs used the deep, feedforward, fully connected neural network architecture. The sizes of the input layers were equal to the length of the subsequence (60). Classification and regression NNs had two hidden layers consisting of 100, 200 (tanh) and 70, 150 (relu) neurons respectively. Between the neuron layers, dropout layers were placed, which randomly shut down of 20% of neurons. The last layer of the classification NN had 3 neurons (softmax) and 1 neuron (tanh) in case of the regression NN. The relu, softmax, tanh are activation functions of neurons. The weights in each layer were optimized using a backward propagation scheme. The Adam optimization algorithm was chosen [17]. NN's were trained on about 500 labelled measurements of χ'(T). The learning of NN's was performed by several dozen epochs. Validation set consisted of about 20% of the dataset. The performance of NN's was tested on 9 hand-selected measurements (test set), which weren't used for learning.

## 5. Results

Both NNs performed well on their task on the prepared dataset. Table 1 presents the values of precision, recall, F1-score parameters. Table 2 shows the confusion matrix. These values were obtained on training data for subsequence length of 60. Figure 6 shows that the predicted value of *d* parameter by regression NN is very close to a ground truth value.

**Table 1.** Precision, recall and F1-score for subsequence length of 60 values.

|         | Precision | Recall | F1-score |
|---------|-----------|--------|----------|
| Class A | 0.99      | 0.99   | 0.99     |
| Class B | 0.96      | 0.85   | 0.90     |
| Class C | 0.93      | 0.98   | 0.95     |
| Mean    | 0.97      | 0.97   | 0.97     |

**Table 2.** Confusion matrix for subsequence length of 60 values.

|                  |         | Real labels |         |         |
|------------------|---------|-------------|---------|---------|
|                  |         | Class B     | Class A | Class C |
| Predicted labels | Class B | 4459        | 242     | 146     |
|                  | Class A | 226         | 24841   | 24      |
|                  | Class C | 517         | 2       | 6972    |

**Table 3.** The predicted $T_c$ by NNs vs value estimated by an expert for the test dataset.

| Sample No. and type of superconductor | Microstructure | $T_c$ by expert (ground truth) [K] | $T_c$ from median value [K] | mean $T_c$ from all subsequences [K] | mean $T_c$ from 3 most probable subsesqences [K] |
|---|---|---|---|---|---|
| 1. $YBa_2Cu_3O_x$ | polycrystalline | 91.51 | 91.59 | 92.11 | 91.72 |
| 2. $YBa_2Cu_3O_x$ | polycrystalline | 91.81 | 91.66 | 91.63 | 91.6 |
| 3. $YBa_2Cu_3O_x$ | polycrystalline | 91.91 | 91.83 | 91.78 | 91.86 |
| 4. $YBa_2Cu_3O_x$ | polycrystalline | 91.49 | 92.15 | 92.19 | 92.36 |
| 5. $YBa_2Cu_3O_x$ | polycrystalline | 91.65 | 91.79 | 91.80 | 91.72 |
| 6. BISCO 2223 | polycrystalline | 106.73 | 107.27 | 107.30 | 107.22 |
| 7. BISCO 2223, | polycrystalline | 107.69 | 107.85 | 107.86 | 107.46 |
| 8. $REBa_2Cu_3O_x$ 2G HTS wire (Superpower) | thin layer | 93.20 | 93.28 | 94.42 | 93.20 |
| 9. $Tl_2Ba_2Ca_2Cu_3O_x$ | polycrystalline | 112.33 | 110.88 | 110.47 | 111.47 |

The final test of the proposed solution was the estimation of $T_c$ on 9 measurements, which were not used for learning the NNs. The results are gathered in Table 3. Selected examples are shown in figure 7. Note that, the χ'(T) curve for sample no. 4 (figure 7b) is noticeably shifted toward negative values, but this obstacle did not confuse the NN from giving a reasonable estimation of $T_c$ value.

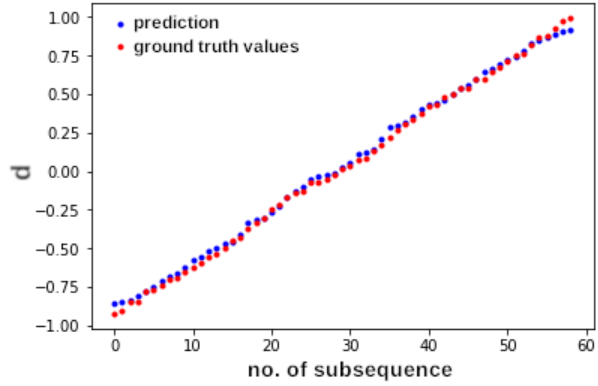

**Figure 6.** Comparison of the estimated value of d parameter vs ground truth for randomly selected measurement.

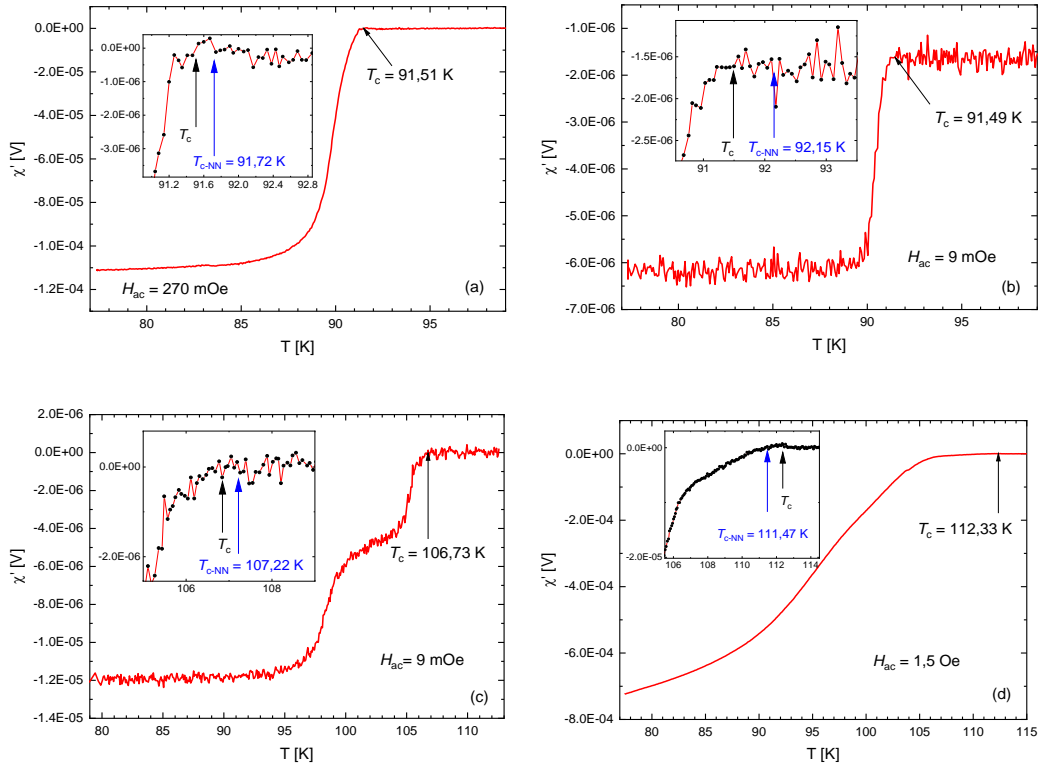

**Figure 7.** A comparison of critical temperatures estimated by NN (marked as $T_{c-NN}$) and by human ($T_c$) for YBa$_2$Cu$_3$O$_x$ - test sample no. 1 (a), YBa$_2$Cu$_3$O$_x$ - no. 4 (b), BISCO 2223 - no. 6 (c), and Tl$_2$Ba$_2$Ca$_2$Cu$_3$O$_x$ no. 9 (d).

## 6. Conclusions

The classification NN was able to recognize with high precision, recall and F1 score parameters whether a generated subsequence of $\chi'(T)$ curve can be used for $T_c$ estimation. The regression NN was able to estimate the value of $T_c$ as well as an expert.

This paper shows proof of concept of working solution and therefore does not focus on finding the most optimal values of parameters like NN architecture. The different length of subsequences 30, 45 etc., can also deliver good results, but a change in subsequence length impose changes in NN architecture (number of neurons) to avoid overfitting of the data. The data issues like uneven distance in temperature between succeeding data points, a different number of data points per measurement should be addressed if one wants to find the best working model.

The proposed approach should also work in tasks of automatic estimation of critical temperatures $T_{c0}$, $T_{onset}$ and finally width of superconducting transition $\Delta T$ from resistance vs temperature measurements. Also, a peak position at maximum value of χ"(T), which is often present in absorption part of AC susceptibility, could be estimated in order to improve critical current determination using Bean model.

The obtained results have its importance in implementation of field induced Josephson junctions made by placement of ferromagnetic strip on the top of superconducting strip or in the construction of Random Access Memory in Rapid Single Quantum Flux computer as described by [18,19].

## Acknowledgements

This work is supported by The National Science Center Poland under Project No. 2018/02/X/ST3/01741. The authors would like to acknowledge and thank Prof. Andrzej Kołodziejczyk from AGH University of Science and Technology and Prof. Janusz Starzyk from University of Information Technology and Management in Rzeszow.